\documentclass[12pt]{article}
\usepackage{latexsym}
\usepackage{amsmath,amssymb}
\usepackage{bm}
\usepackage[dvipdfmx]{graphicx,color}
\usepackage{amssymb}
\usepackage{cite}
\usepackage{amsfonts, amscd, amsthm}
\usepackage[all]{xy}
\usepackage{here}
\usepackage{ulem}
\usepackage{mathabx}

\renewcommand{\theequation}{\thesection.\arabic{equation}}

\newcommand{\dd}{{\rm d}}

\newcommand{\x}{{\bf x}}
\newcommand{\B}{{\bf B}}
\newcommand{\A}{{\bf A}}
\newcommand{\F}{{\bf F}}
\newcommand{\E}{{\bf E}}
\newcommand{\p}{{\bf p}}

\makeatletter
\@addtoreset{equation}{section}
\renewcommand{\theequation}{\thesection.\@arabic\c@equation}
\makeatother

\makeatletter
\renewcommand\appendix{\par
  \setcounter{section}{0}%
  \setcounter{subsection}{0}%
  \gdef\thesection{Appendix \@Alph\c@section }
  \renewcommand{\theequation}
  {\Alph{section}.\arabic{equation}}
}
\makeatother

\makeatletter
\newcounter{subeqncnt}
\def\thesubeqncnt{\alph{subeqncnt}}
\def\subequations{\begingroup%
\stepcounter{equation}\edef\@tempa{\theequation}%
\let\c@equation\c@subeqncnt\c@subeqncnt\z@
\edef\theequation{\@tempa\noexpand\thesubeqncnt}}

\makeatother

\allowdisplaybreaks

\begin{document}

\titlepage

\title{Comments on the Aharonov-Bohm Effect} 


\author{
Kazuyasu Shigemoto and Kunihiko Uehara \\
Tezukayama University, Nara 631-8501, Japan\\
}
\date{\empty}


\maketitle
\abstract{
In the original setting of the Aharonov-Bohm, the gauge invariant 
physical longitudinal mode of the vector potential, 
which is written by the gauge invariant physical current 
$(-e)\bar{\psi}  {\boldsymbol \gamma} \psi$, gives the desired contribution to the  
Aharonov-Bohm effect. While the scalar 
mode of the vector potential, which changes under the gauge transformation 
so that it is the unphysical mode, give no contribution to the Aharonov-Bohm effect. 
Then Aharonov-Bohm effect really occurs by the physical longitudinal mode
in the original Aharonov-Bohm's setting.
In the setting of Tonomura {\it et al.}, where the magnet is shielded with the 
superconducting material, not only the magnetic field but also the longitudinal 
mode of the vector potential become massive by the Meissner effect.
Then not only the magnetic field but also the physical
longitudinal mode does not
come out to the region where the electron travels. In such setting, only the 
scalar mode of the vector potential exists in the region where the electron
travels, but there is no contribution to  the 
Aharonov-Bohm effect from that mode. Then, theoretically, the 
Aharonov-Bohm effect does not occur in the Tonomura {\it et al.}'s setting.

In the quantum theory, the electron is treated as the wave, and the longitudinal
mode give the change of the phase, which gives the Aharonov-Bohm effect.
In the classical theory, the electron is treated as the particle, and the only 
existing longitudinal 
mode gives the change of the angular momentum. For the particle, there is no
concept of the phase, so that there is no Aharonov-Bohm effect.
}

\section{Introduction} 
\setcounter{equation}{0}

In 1959, Aharonov and Bohm publish the paper on the issue of the vector 
potential, which changes under the gauge transformation, so that the vector potential 
is unphysical,  causes the 
physical observable Aharonov-Bohm effect\cite{Aharonov1,Aharonov2,Aharonov3}. 
There is another issue that the Aharonov-Bohm effect is the genuine quantum 
effect or not\cite{Aharonov1,Aharonov2,Aharonov3}. This is because, in the classical
theory, there does not give any force through the Lorentz force with zero magnetic field
 in the region where the electron travels. 
 
After their paper, various experimental test of the Aharonov-Bohm effect
were reported~\cite{Chambers,Fowler,Mollenstedt,Boersch}.  There also came various 
theoretical explanations different from that of Aharonov-Bohm, that is, 
the explanation by the non-local field instead of the local 
vector potential~\cite{Noerdlinger,DeWitt,Belinfante}, the explanation 
by a new force 
in addition to the Lorentz force~\cite{Liebowitz}, the explanation by including the 
back-reaction of the solenoid~\cite{Boyer1,Boyer2} etc..

Then there has come the theoretical paper that the Aharonov-Bohm effect itself 
does not occur~\cite{Bocchieri1,Bocchieri2}.
Furthermore there have appeared various theoretical and experimental papers to discuss 
whether the Aharonov-Bohm effect really exists or not and  
whether the Aharonov-Bohm effect is the genuine quantum effect or 
not~\cite{Bohm,Klein,Olariu}.

In order to clarify such situations, Tonomura {\it et al.} proposed 
the experiment to shield a magnet with the superconducting material~\cite{Tonomura1,Tonomura2}, 
because the main objection of the non-existence of the Aharonov-Bohm
effect is that the experimental evidence comes from the leakage of the 
magnetic field.
  
In order to clarify the argument of the Aharonov-Bohm effect, we consider 
the Aharonov-Bohm effect caused by the infinitely long solenoid 
with the steady current.

In order to make the argument clear, 
we physically decompose the vector potential $\A$
into various modes as the solution of the Maxwell equation.  Time-dependent modes are two transverse modes $\A^{T}$, 
which become the photon when quantized.   The rests are the longitudinal 
mode $\A^{L}$ which is 
expressed by the integral of the current, and the scalar 
mode $\A^{S}$ which is the pure gauge mode. 
As we focus on the Aharonov-Bohm effect, 
we consider only the longitudinal mode $\A^{L} (\x)$ and the scalar mode
$\A^{S} (\x)$ of the vector potential $\A (\x)$. 

The longitudinal mode $\A^{L}(\x)$ is the projected mode from $\A(\x)$ in the momentum 
space in the form $\A^{L}(\x)_i=(\delta_{i j}-\partial_i \frac{1}{\triangle}\partial_j)\A(\x)_j
=P_{ij}\A(\x)_j$ with $P_{ij}P_{jk}=P_{ik}$, 
and the longitudinal mode is uniquely given as the solution of the Maxwell equation 
in the form
\begin{equation}
\A^{L}(\x, t)=\frac{\mu_0}{4 \pi} \int \dd^3 x' \frac{ {\bf j}(\x ', t)}{|\x-\x '|} .
\label{1e1}
\end{equation}
For the electric current, we have the explicit form 
${\bf j}(\x, t)=(-e) \bar{\psi}(\x,t) {\boldsymbol \gamma} \psi(\x, t)$ by using the 
electron field $\psi(\x, t)$. 
Though the phase 
of the electron changes, the current does not change under the 
gauge transformation, which means that the longitudinal mode does not 
change under the gauge transformation.
The scalar mode $\A^{S} (\x)$ is the pure gauge mode of the form 
\begin{equation}
\A^{S}(\x, t)=\nabla \theta(\x,t) ,
\label{1e2}
\end{equation}
where $\theta(\x,t)$ is the scalar function for the gauge transformation. 
We call the physical mode which does not change, and the unphysical mode 
which changes, under the gauge transformation.
Then the transverse and the longitudinal modes are physical mode and the scalar
mode is the unphysical mode.

\section{The Aharonov-Bohm effect in various settings}
\setcounter{equation}{0}
\subsection{The original Aharonov-Bohm setting}

We review the original Aharonov and Bohm setting, and consider the effect from the
infinitely long solenoid.

For the infinite long solenoid with the radius  $a$ of the form 
$\rho=a$, $\varphi=[0, 2 \pi], z=[-\infty, +\infty]$ in the cylindrical coordinate 
$x=\rho \cos \varphi$, $y=\rho \sin\varphi$, $z=z$ and the longitudinal mode 
of the vector potential outside the solenoid is given by
\begin{equation}
\A^{L}(\x)_{\rho}=0, \quad \A^{L}(\x)_{z}=0, \quad 
\A^{L}(\x)_{\varphi}=\frac{\mu_0 n I a^2}{2 \rho}  , 
\label{2e1}
\end{equation}
where $I$ is the steady current, $n$ is the number of turns of the solenoid in the unit length 
for the $z$-direction.
The contribution to the magnetic field  $\B^{L}(\x)$
from the longitudinal mode of the vector potential $\A^{L}(\x)$ is given to be zero 
in the following way
\begin{eqnarray}
&& \B^{L}_{\rho}=\frac{1}{\rho} \frac{\partial \A^{L}_{z}}{\partial \varphi}
-\frac{\partial \A^{L}_{\varphi}}{\partial z} =0, \quad
\B^{L}_{\varphi}=\frac{\partial \A^{L}_{\rho}}{\partial z}
-\frac{\partial \A^{L}_{z}}{\partial \rho} =0   ,  
\label{2e2}\\
&&\B^{L}_{z}=\frac{1}{\rho} \frac{\partial (\rho \A^{L}_{\varphi})}{\partial \rho}
-\frac{1}{\rho} \frac{\partial \A^{L}_{\rho}}{\partial \varphi} 
=\frac{1}{\rho}\frac{\partial  (\mu_0 n I a^2/2)}{\partial \rho}=0  .
\label{2e3}
\end{eqnarray}
The contribution to the magnetic field $\B^{S}(\x)$ from the scalar mode of the
vector potential $\A^{S}(\x)=\nabla \theta(\x) $ is trivially given to be zero.
In this way, we confirm that there is no magnetic field $\B^{S}(\x)=0$, $\B^{L}(\x)=0$,
around the region where the electron travels.
While the line integral of the longitudinal mode around the closed curve $C$ gives 
\begin{eqnarray}
\oint_C  \A^{L}(\x) \cdot \dd \x 
=\int_{0}^{2 \pi}  \A^{L}(\x)_{\varphi}\ \rho\ \dd \varphi
=\frac{\mu_0 n I a^2}{2} \int_{0}^{2 \pi} \dd \varphi 
=\mu_0 n I a^2 \pi  .
\label{2e4}
\end{eqnarray}
We have the same result by using the Stokes' theorem.

While the line integral of the scalar mode around the closed curve $C$ gives 
\begin{eqnarray}
\oint_C  \A^{S}(\x) \cdot \dd \x =\oint_C  \nabla \theta(\x) \cdot \dd \x 
=\oint_C  {\rm d} \theta(\x) =\theta({\rm P})-\theta({\rm P} )=0  ,
\label{2e5}
\end{eqnarray}
where the closed curve $C$ starts from the point P and ends at the same point P.
Then, in the quantum theory, the contribution from only the physical mode, the longitudinal mode, 
gives the change of the phase of the electron  
$\displaystyle{\Phi=\frac{(-e)}{\hbar} \oint \A^{L}(\x) \cdot \dd \x
=\frac{(-e) \mu_0 nI a^2 \pi}{\hbar}}$, which gives the Aharonov-Bohm effect.

In the classical theory, as the Lorentz force to an electron is given by
\begin{equation}
\F=(-e) (\E+{\bf v} \times \B)=m {\bf a}  , 
\label{2e6}
\end{equation}
and $\phi(\x)=0$, $\B(\x)=\B^{L}(\x)+\B^{S}(\x)=0$ in the Aharonov-Bohm setting, we have
\begin{equation}
\F=-(-e) \frac{\dd \A^{L}(\x(t))}{\dd t}=m \frac{\dd {\bf v}(t)}{\dd t} , 
\label{2e7}
\end{equation}
where we use the longitudinal mode of Eq.(\ref{2e1}) with the case of 
a very slow changing current, which becomes the steady current in the final
limit.
Then, classically,  we have the conservation of the total momentum
\begin{equation}
\p(t)=m {\bf v}(t)+(-e) \A^{L}(\x(t)) =\text{(const.)}  .
\label{2e8}
\end{equation}
We denote $\p^{L}(t)=(-e) \A^{L}(\x(t))$ as the momentum of the longitudinal mode 
of the vector potential. Classically, as the electron is treated as the particle, we 
cannot include the scalar mode 
$\A^{S}(\x, t)=\nabla \theta(\x, t)$ in the momentum
because there is no concept of the phase 
factor of the electron. (In the quantum theory, the scalar mode and the gradient of the phase factor 
of the electron, which change under the gauge transformation,  
compensate to give the theory  
gauge invariant.) Classically, there is no concept of the phase of the electron, there is not Aharonov-Bohm
effect, but the physically meaningful longitudinal mode gives the change of the 
angular momentum
$\displaystyle{\oint \p^{L}(\x) \cdot \dd \x=(-e) \oint \A^{L}(\x) \cdot \dd \x=(-e) \mu_0 nI a^2 \pi}$.
As is well-known, 
$\displaystyle{\oint  \p(\x) \cdot \dd \x}=nh$ gives the Bohr's semi-classical
quantization condition.
In this way, in the classical level, the physically meaningful longitudinal mode of the 
vector potential gives the change of the angular momentum.

On the issue that the vector potential, which changes under the gauge 
transformation, gives the physical Aharonov-Bohm effect, we have the
following conclusion.
In the quantum theory, the corresponding vector potential is composed of the longitudinal and the 
scalar modes.  The longitudinal mode, which does not change under the gauge 
transformation,  gives the observed Aharonov-Bohm effect. 
The scalar mode, which changes under the gauge transformation, gives no
contribution to the Aharonov-Bohm effect.

In the classical theory, there does not appear the scalar mode in the momentum.
Thus the physical longitudinal mode only contributes the physical change of the angular momentum.

.

\subsection{The Tonomura {\it et al.}'s setting}
In this case, we consider the setting that the infinitely long solenoid is shielded by 
the superconducting material.
The theoretical model is known as the Ginzburg-Landau-Abrikosov-Gor'kov 
theory~\cite{Ginzburg,Abrikosov} in the form 
\begin{eqnarray}
\frac{1}{2m}\big|\left(-i \hbar \nabla -(-e^*) \right) \A^{L}\big|^2 \psi+\alpha \psi+\beta |\psi|^2 \psi=0,
\label{2e9}
\end{eqnarray}
where $(-e^{*})$ is the effective charge for the Cooper pair with $(-e^{*})=(-2 e)$, 
which explains the Meissner effect~\cite{Meissner} and the London equation~\cite{London}. 
The Ginzburg-Landau paper is the origin of the Higgs mechanism. 
In the Tonomura {\it et al.}'s setting, the longitudinal mode of the vector potential becomes 
massive through 
the Higgs mechanism, which is known as the Meissner effect, that is, 
$|\A^{L}(\x)| \sim \exp(- |\x|/\lambda_p)$  as $|\x| \rightarrow \infty$, with the 
penetration length $\lambda_p=\sqrt{m/4\mu_0 (e^*)^2 |\psi(\infty)|}$.
Then, not only the magnetic field but also the longitudinal mode of the vector potential
well decays at a long distance. So that not only the magnetic field but also 
the longitudinal mode does not exist, that is, $\B(\x)=0$, $\A^{L}(\x)=0$
in the region where the electron travels. 
Idealistically,  we consider that there is the forbidden region, and inside 
the forbidden region, where the magnetic field and the longitudinal mode of the 
vector potential exist. The electron travels in the allowed 
region.  In this allowed region, $\B(\x)=0$, $\A^{L}(\x) =0$, (neither leakage of the 
magnetic field nor the longitudinal mode of the vector potential). While the scalar 
mode is possible to exist $\A^{S}(\x)=\nabla\theta(\x) \ne 0$, 
but as we have already shown, the scalar mode trivially does not contribute
to the Aharonov-Bohm effect.

In this Tonomura {\it et al.}'s setting, when we apply the Stokes' theorem, we must  
carefully use the theorem for
the non-simply connected region in the form
\begin{equation}
\iint_{D} {\bf B}_n \cdot \dd S =\oint_{C_1} {\bf A} \cdot \dd {\bf x}
+\oint_{C_2} {\bf A} \cdot \dd {\bf x} ,
\label{2e10}
\end{equation}
where $\B=\nabla \times \A$ and $C_1$, $C_2$ and $D$ are shown in Figure 1.
%
%
%
\begin{figure}[H]
\begin{center}
\includegraphics[width=50mm]{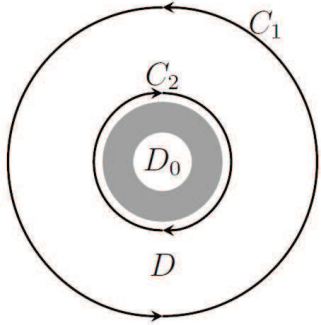}
\end{center}
\caption{Two closed curves and the region $D$.}
\label{Fig1}
\end{figure}
We give the explanation of the generalized Stokes' theorem in the Appendix.

In Tonomura {\it et al.}'s setting of the Aharonov-Bohm effect, we have 
$\A(\x)=\A^{S}(\x)=\nabla \theta(\x)$,
and we have 
\begin{eqnarray}
&&\oint_{C_1} \A^{S} \cdot \dd \x=\oint_{C_1} \nabla \theta(\x) \cdot \dd \x
=\oint_{C_1} \dd \theta(\x)=0 ,
\label{2e11}\\
&&\oint_{C_2} \A^{S} \cdot \dd \x=\oint_{C_2} \dd \theta(\x)=0 .
\label{2e12}
\end{eqnarray}
While, as there is no magnetic field in the region $D$, we have
\begin{equation}
\iint_{D} {\bf B}_n \cdot \dd S =0  .
\label{2e13}
\end{equation}
Then the Stokes' theorem of Eq.(\ref{2e10}) is trivially satisfied by using relations
\begin{equation}
\iint_{D} {\bf B}_n \cdot \dd S=0 , \quad
\oint_{C_1} \A^{S} \cdot \dd \x=0, \quad
\oint_{C_2} \A^{S} \cdot \dd \x=0.  
\label{2e14}
\end{equation}
In the Stokes' theorem in the Tonomura {\it et al.}'s setting, we must
notice the following fact
\begin{equation}
\iint_{D_0+D} {\bf B}_n \cdot \dd S 
\ne \oint_{C_1} {\bf A} \cdot \dd {\bf x}  .
\label{2e15}
\end{equation}
In the Tonomura {\it et al.}'s setting, we have $\A(\x)=\A^{S}(\x)=\nabla \theta(\x)$
in $D$, then the right-hand side of Eq.(\ref{2e15}) gives zero. While the magnetic 
field inside the forbidden region $D_0$
gives the contribution to the left-hand side
of the value $\mu_0 n I a^2 \pi$. Therefore, the left-hand side  is different from 
the right-hand side of Eq.(\ref{2e15}).

In order to avoid the mistake, though it may be mathematically 
complicated, it will be better to calculate the phase shift only 
by using the vector potential, but not by using the magnetic 
field with the help of the Stokes' theorem.
\newpage

\section{Summary and Discussions} 
\setcounter{equation}{0}

We clarify issues of the Aharonov-Bohm effect. 
In the original setting of the Aharonov-Bohm, the longitudinal mode 
of the vector potential, which does not change under the gauge 
transformation so that it is the physical mode, gives the contribution of the observed
Aharonov-Bohm effect. While the scalar 
mode of the vector potential, which changes under the gauge transformation 
so that it is the unphysical mode, gives no contribution to the Aharonov-Bohm effect. 
Then Aharonov-Bohm effect really occurs by the physical longitudinal mode
in the original Aharonov-Bohm setting.
In the setting of Tonomura {\it et al.}, where the magnet is shielded by the 
superconducting material, not only the magnetic field but also the longitudinal 
mode of the vector potential become massive by the Meissner effect.
Then not only the magnetic field but also the longitudinal mode does not
come out to the region where the electron travels. In such setting, only the 
scalar mode of the vector potential exists, but the contribution to  the 
Aharonov-Bohm effect from that mode is zero. Then, theoretically, the 
Aharonov-Bohm effect does not occur for Tonomura {\it et al.}'s setting.
As people misuse the 
Stokes' theorem for the non-simply connected region, we explain the 
correct Stokes' theorem for the non-simply connected region.

In the quantum theory, the electron is treated as the wave, and the longitudinal
mode gives the change of the phase, which causes the Aharonov-Bohm effect.
In the classical theory, the electron is treated as the particle, and the longitudinal 
mode gives the change of the angular momentum. For the particle, there is no
concept of the phase, so that  there is no Aharonov-Bohm effect.

\vskip 20mm


\appendix
\section{The Stokes' theorem in the non-simply connected region} 
\setcounter{equation}{0}

The Stokes' theorem in the non-simply connected region of Figure 2  is given 
in the form
\begin{equation}
\iint_{D} {\bf B}_n \cdot \dd S =\oint_{C_1} {\bf A} \cdot \dd {\bf x}
+\oint_{C_2} {\bf A} \cdot \dd {\bf x} ,
\label{3e1}
\end{equation}
where $ {\bf B}= \nabla \times \A$.
%
%
\begin{figure}[H]
\begin{center}
\includegraphics[width=50mm]{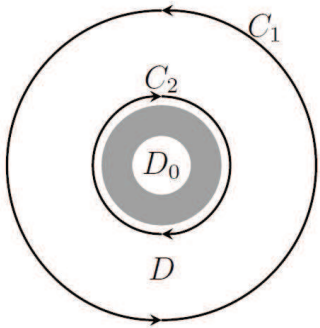}
\end{center}
\caption{Two closed curves and the region $D$.}
\label{Fig2}
\end{figure}

We explain the Stokes'  theorem in the non-simply connected region in the following.
We start from 
\begin{equation}
\oint_{C_1} {\bf A} \cdot \dd {\bf x}
+\oint_{C_2} {\bf A} \cdot \dd {\bf x}  ,
\label{3e2}
\end{equation}
which is rewritten into  the infinitely many line integrals around closed curves as in Figure 3,
which is equivalent to the sum of the magnetic field perpendicular 
to the surface $\displaystyle{\iint_{D} {\bf B}_n \cdot \dd S }$.
%
%
\begin{figure}[H]
\begin{center}
\includegraphics[width=50mm]{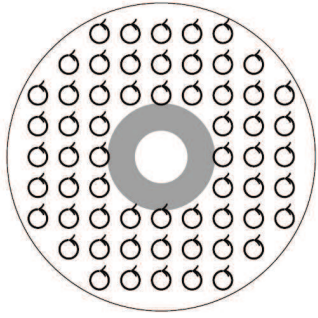}
\end{center}
\caption{Infinitely many closed curves in the region $D$. }
\label{Fig3}
\end{figure}
While the two line integrals around closed curves $C_1$ and $C_2$ in Figure 2 can 
be transformed into the single line integral of the closed curve 
$C=C_1+C_4+C_2+C_3$ as is given in Figure 4.
\begin{equation}
\oint_{C_1} {\bf A} \cdot \dd {\bf x}
+\oint_{C_2} {\bf A} \cdot \dd {\bf x}
=\oint_{C_1} {\bf A} \cdot \dd {\bf x}
+\oint_{C_4} {\bf A} \cdot \dd {\bf x}
+\oint_{C_2} {\bf A} \cdot \dd {\bf x}
+\oint_{C_3} {\bf A} \cdot \dd {\bf x}
=\oint_{C} {\bf A} \cdot \dd {\bf x}  , 
\label{3e3}
\end{equation}
where $C=C_1+C_4+C_2+C_3$ is the single closed curve,
which does not enclose the forbidden region $D_0$. 
In this way, even for the non-simply connected region, the Stokes' theorem
is the theorem of the single line integral of the closed curve, which does 
not enclose the forbidden region.  We can topologically shrink this closed line 
integral around the closed curve $C$ into the infinitesimal closed line integral 
of the closed curve around any one point in the region $D$. 

%
%
\begin{figure}[H]
\begin{center}
\includegraphics[width=50mm]{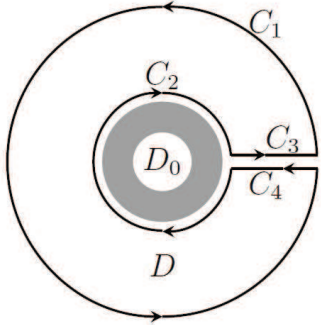}
\end{center}
\caption{Single closed curve $C=C_1+C_4+C_2+C_3$ 
around the region $D$. }
\label{Fig4}
\end{figure}

For the Stokes' theorem applied to the general 
non-simply connected region, we must take the single closed curve 
$C$ in such a way as the closed curve can be topologically transformed to 
the infinitesimal closed curve around any one point in the region $D$.

\newpage


\begin{thebibliography}{99}
\bibitem{Aharonov1}
Y. Aharonov and D. Bohm, 
``Significance of Electromagnetic Potentials in the Quantum Theory",
Phys. Rev.  {\bf 115}, 485-491 (1959). 
%
\bibitem{Aharonov2}
Y. Aharonov and D. Bohm, 
``Further Considerations on Electromagnetic Potentials in the Quantum Theory",
Phys. Rev.  {\bf 123}, 1511-1524 (1961). 
%
%
\bibitem{Aharonov3}
Y. Aharonov and D. Bohm, 
``Remarks on the Possibility of Quantum Electrodynamics without Potentials",
Phys. Rev.  {\bf 125}, 2192-2193 (1962). 
%
%
\bibitem{Chambers}
R.G. Chambers, 
``Shift of an Electron  Interference Pattern by Enclosed Magnetic Flux",
Phys. Rev. Lett.  {\bf 5}, 3-5 (1960). 
%
%
\bibitem{Fowler}
H.A. Fowler, L. Marton, J. A. Simpson and J. A. Suddeth
``Electron Interferometer Studies of Iron Whiskers",
J. Appl. Phys. {\bf 32}, 1153-1155 (1961). 
%
%
\bibitem{Mollenstedt}
G. M\"{o}llenstedt and W. Bayh, 
``Messung der kontinuierlichen Phasenschiebung von Elektronenwellen 
im kraftfeldfreien Raum durch das magnetische vektorpotential einer Luftspule",
Naturwissenschaften {\bf 49}, 81-82 (1962). 
%
%
\bibitem{Boersch}
H. Boersch, H. Hamisch, K. Grohmann and D. Wohlleben 
``Experimenteller Nachweis der Phasenschiebung von Elektronenwellen 
durch das magnetische Vektorpotential",
Z. Phys.  {\bf 165}, 79-93 (1961). 
%
%
\bibitem{Noerdlinger} 
P.D. Noerdlinger,  
``Elimination of the Electromagnetic Potentials",
Nuovo Cimento {\bf 23}, 158-167 (1962).
%
%
\bibitem{DeWitt} 
B.S. DeWitt,  
``Quantum Theory without Electromagnetic Potentials",
Phys. Rev. {\bf 125}, 2189-2191 (1962).
%
%
\bibitem{Belinfante} 
F.J. Belinfante,  
``Consequences of the Postulate of a Complete Commuting Set of
Observables in Quantum Electrodynamics",
Phys. Rev. {\bf 128}, 2832-2837 (1962).
%
%
\bibitem{Liebowitz} 
B. Liebowitz,  
``Significance of the Aharonov-Bohm Effect.",
Nuovo Cimento {\bf 38}, 932-950 (1965).
%

%
\bibitem{Boyer1} 
T.H. Boyer,  
``Classical Electromagnetic Interaction of a Charged
Particle with a Constant-current Solenoid",
Phys. Rev. {\bf D 8}, 1679-1678 (1973).
%
%
\bibitem{Boyer2} 
T.H. Boyer,  
``Classical Electromagnetic Deflections and Lag Effects Associated
with Quantum Interference Pattern Shifts: Considerations
Related to the Aharonov-Bohm Effect",
Phys. Rev. {\bf D 8}, 1667-1693 (1973).
%
%
\bibitem{Bocchieri1} 
P. Bocchieri and A. Loinger,  
``Nonexistence of the Aharonov-Bohm Effect",
Nuovo Cimento {\bf 47A}, 475-482 (1978).
%
\bibitem{Bocchieri2} 
P. Bocchieri, A. Loinger and G. Siragusa,  
``Nonexistence of the Aharonov-Bohm Effect II",
Nuovo Cimento {\bf 51A}, 1-17 (1979).
%
%
\bibitem{Bohm} 
D. Bohm and B.J. Hiley,  
``On the Aharonov-Bohm Effect",
Nuovo Cimento {\bf 52A}, 295-308 (1979).
%
%
\bibitem{Klein} 
U. Klein,  
``The inadmissibility of non-Stokesian vector potentials in quantum mechanics. 
Comments on a paper asserting the nonexistence of the Aharonov-Bohm effect",
Lett. Nuovo Cimento {\bf 25}, 33-37 (1979).
%
%
\bibitem{Olariu} 
S. Olariu and I.I. Popescu,  
``The quantum effects of electromagnetic fluxes",
Rev. Mod. Phys. {\bf 57}, 339-436 (1985).
%
%
%
\bibitem{Tonomura1}
A. Tonomura, N. Osakabe, T. Matsuda, T. Kawasaki, J. Endo, 
S. Yano, and H. Yamada, 
``Evidence for Aharonov-Bohm Effect with Magnetic Field Completely 
Shielded from Electron Wave", 
Phys. Rev. Lett. {\bf 56}, 792-795 (1986).
%
%
\bibitem{Tonomura2}
N. Osakabe, T. Matsuda, T. Kawasaki, J. Endo, A. Tonomura, S. Yano and H. Yamada, 
``Experimental Confirmation of Aharonov-Bohm Effect using a Toroidal 
Magnetic Field Confined by a Superconductor", 
Phys. Rev. {\bf A 34}, 815-822 (1986).
%
%
\bibitem{Ginzburg}
V.L. Ginzburg and L.D. Landau,  
``On the Theory of Superconductivity", 
Zh. Eksp. Teor. Fiz. {\bf 20}, 1064-1082 (1950).
%
\bibitem{Abrikosov}
A.A. Abrikosov and L.P. Gor'kov,  
``Contribution to the Theory of Superconducting Alloys with Paramagnetic Impurities", 
Sov. Phys. JETP {\bf 12}, 1243-1253 (1961).
%
%
\bibitem{Meissner}
W. Meissner and R. Ochsenfeld,  
`` Ein neuer Effekt bei Eintritt der Supraleitfähigkeit",  
 Naturwissenschaften {\bf 21}, 787-788 (1933).
%
%
\bibitem{London}
F. London and H. London,  
`` The Electromagnetic Equations of the Supraconductor", 
Proc. Roy. Soc. {\bf A 149}, 71-88 (1935).
%
\end{thebibliography}
\end{document}